\documentclass[12pt,reqno]{amsart}
\usepackage{amsmath,amsfonts,amsthm,amssymb}

\newtheorem{thm}{Theorem}
\newtheorem{conj}[thm]{Conjecture}
\newtheorem{lem}{Lemma}[section]
\newtheorem{cor}[thm]{Corollary}
\newtheorem{prop}[lem]{Proposition}

\newcommand{\eq}[1]{eq.~(\ref{#1})}  
\newcommand{\norm}[1]{\left\Vert#1\right\Vert}
\newcommand{\abs}[1]{\left\vert#1\right\vert}
\newcommand{\set}[1]{\left\{#1\right\}}
\newcommand{\com}[2]{\left[ #1 , #2 \right ]}
\newcommand{\ip}[2]{\left < #1, #2 \right >}

\def\eps{\varepsilon}

\def\e{\mathrm e}
\def\im{\mathrm i}
\def\Im{\mathrm{Im}}
\def\half {\frac{1}{2}}
\def\1{{\mathsf 1}}
\def\di{\mathrm d}
\def\din{\mathrm d}
\def\diq{ \frac{\di\mathbf{q}}{(2 \pi)^d}}
\def\dik{\frac{\di\mathbf{k}}{(2 \pi)^d}}

\def\Z{\mathbb Z}

\def\R{\mathbb R}

\def\E{\mathbb E}

\def\Pr{\mathbb{P}} 
\def\esssup{\operatorname*{ess-sup}}

\begin{document}

\title[Density of states at weak disorder]{H\"older equicontinuity of
the integrated density of states at weak disorder}
\author{Jeffrey H Schenker}
\date{31 March 2004, revised 14 July 2004}
\address{Institut f\"ur Theoretische Physik \\ ETH, Z\"urich \\ CH-8093 Z\"urich}
\email{jschenker@itp.phys.ethz.ch} 


\begin{abstract}
H\"older continuity, $|N_\lambda(E)-N_\lambda(E')|\le C |E-E'|^\alpha$, with a
constant $C$ independent of the disorder strength $\lambda$ is proved for the
integrated density of states $N_\lambda(E)$ associated to a discrete random
operator $H = H_o + \lambda V$ consisting of a translation invariant hopping
matrix $H_o$ and i.i.d. single site potentials $V$ with an absolutely
continuous distribution, under a regularity assumption for the hopping term.
\end{abstract}
\maketitle
\section{Introduction}

Random operators on $\ell^2(\Z^d)$ of the general form
\begin{equation}\label{eq:mainequation}
    H_\omega \ = \ H_o + \lambda V_\omega \; ,
\end{equation}
play a central role in the theory of disordered materials, where:
\begin{enumerate}
  \item $V_\omega
  \psi(x)  =  \omega(x) \psi(x)$ with $\omega(x)$, $x \in \Z^d$, independent
  identically distributed random variables
  whose common distribution is
  $\rho(\omega) \di \omega$ with $\rho$ a bounded function.
  The coupling $\lambda \in \R$ is called the {\em disorder
  strength}.
  \item $H_o$ is a bounded translation invariant operator, i.e.,
  $\com{S_\xi} {H_o} =0$ for each translation $S_\xi \psi(x) =
  \psi(x-\xi)$, $\xi \in \Z^d$.
\end{enumerate}
The {\em density of states measure} for an operator $H_\omega$ of the form
\eq{eq:mainequation} is the (unique) Borel measure $\di N_\lambda(E)$ on the
real line defined by
\begin{equation*}
    \int f(E) \di N_\lambda (E) \ = \ \lim_{L \rightarrow \infty}
    \frac{1}{\#\set{x \in
    \Z^d : |x| < L}} \sum_{x: |x| < L}
    \ip{\delta_x}{f(H_\omega)\delta_x} \; ,
\end{equation*}
and the {\em integrated density of states} $N_\lambda(E)$ is
\begin{equation*}
    N_\lambda(E) \ := \ \int_{(-\infty,E)} \di N_\lambda(\eps) \; .
\end{equation*}
It is a well known consequence, e.g., ref \cite{PasturFigotin1992}, of the
translation invariance of the distribution of $H_\omega$ that the density of
states exists and equals
\begin{equation*}
    N_\lambda (E) \ = \ \int_\Omega
    \ip{\delta_0}{P_{(-\infty,E)}(H_\omega)\delta_0} \di
    \Pr(\omega)\; , \quad \mbox{ every $E \in \R$};
\end{equation*}
for $\Pr$ almost every $\omega$, where $\Pr$ is the joint probability
distribution for $\omega$ and $\Omega=\R^{\Z^d}$ is the probability space.

The density of states measure is an object of fundamental physical interest.
For example, the free energy $f$ per unit volume of a system of non-interacting
identical Fermions, each governed by a Hamiltonian $H_\omega$ of the form
\eq{eq:mainequation}, is
\begin{equation*}
    f(\mu,\beta) \ = \ - \beta \int \ln (1+ \e^{-\beta(E - \mu)}) \di
    N_\lambda(E) \; ,
\end{equation*}
where $\beta$ is the inverse temperature and $\mu$ is the chemical potential.
Certain other thermodynamic quantities (density, heat capacity, etc.) of the
system can also be expressed in terms of $N_\lambda$.

Our main result is equicontinuity of the family $\{N_\lambda(\cdot),\lambda
>0\}$ within a class of H\"older continuous functions, that is
\begin{equation}\label{eq:mainresult}
      N_\lambda(E+\delta) - N_\lambda(E - \delta)
    \ \le \ C_\alpha \, \delta^\alpha,
    \text{ for all } \lambda > 0 \; ,
\end{equation}
under appropriate hypotheses on $H_o$. The exponent $\alpha <1$ depends on
$H_o$ as well as the probability density, with $\alpha =\half$ at generic $E$
for a large class of hopping terms if $\rho$ is compactly supported.

A bound of the form \eq{eq:mainresult} for the integrated density of states
associated to a continuum random Schr\"odinger operator is implicit in Theorem
1.1 of ref.~\cite{CombesHislopKlopp2003}, although uniformity in $\lambda$ is
not explicitly noted there.  The tools of ref.~\cite{CombesHislopKlopp2003}
carry over easily to the discrete context to give an alternative proof of
\eq{eq:mainresult}. However the methods employed herein are in fact quite
different from those of ref.~\cite{CombesHislopKlopp2003}, and may be
interesting in and of themselves.

The main point of \eq{eq:mainresult} is the uniformity of the
bound as $\lambda \rightarrow 0$, since the well known {\em Wegner
estimate} \cite{Wegner1981}, see also \cite[Theorem
8.2]{Spencer1984},
\begin{equation}\label{eq:Wegner}
    \frac{\di N_\lambda(E)}{\di E} \ \le \ \frac{ \norm{\rho}_\infty}{\lambda}
    \; ,
\end{equation}
implies that $N_\lambda(E)$ is in fact Lipschitz continuous,
\begin{equation}\label{eq:Lipschitz}
    N_\lambda (E+\delta)-N_\lambda(E-\delta) \ \le \ \frac{ \norm{\rho}_\infty}{\lambda} \
    2\delta \; .
\end{equation}
However, the Lipschitz constant $\norm{\rho}_\infty / \lambda$ in
\eq{eq:Lipschitz} diverges as $\lambda \rightarrow 0$.  Such a singularity is
inevitable for a bound which makes no reference to the hopping term, since $
\di N_\lambda(E) = \lambda^{-1} \rho(E/\lambda) \di E $ for $H_o = 0$, as may
easily be verified. However if the background itself has an absolutely
continuous density of states, the Wegner estimate is far from optimal at weak
disorder.

The translation invariant operator $H_o$ may be written as a
superposition of translations,
\begin{equation*}
    H_o \ = \ \sum_{\xi} \check{ \eps}(\xi)S_\xi \; ,
\end{equation*}
where
\begin{equation*}
    \check{\eps}(\xi) \ = \ \int_{T^d} \eps(\mathbf q) \e^{-\im \xi \cdot \mathbf q}
     \, \diq \; ,
\end{equation*}
is the inverse Fourier transform of a bounded real function $\eps$ on the torus
$T^d = [0, 2 \pi)^d$, called the symbol of $H_o$. For any bounded measurable
function $f$,
\begin{equation*}
    f(H_o) \ = \ \sum_{\xi \in \Z^d} \left [  \int_{T^d} f(\eps(\mathbf q))
    \e^{-\im \xi \cdot \mathbf q}
     \, \diq \right ] S_\xi \; ,
\end{equation*}
from which it follows that the density of states $N_o(E)$ for $H_o$ obeys
\begin{equation*}
    \int f(E) \di N_o(E) \ = \ \int_{T^d} f(\eps(\mathbf q))
     \, \diq \; .
\end{equation*}
In particular,
\begin{equation*}
    N_o(E) \ = \ \int_{\set{\eps(\mathbf q) < E}} \diq
    \; .
\end{equation*}

We define a {\em regular point} for $\eps$ to be a point $E \in
\R$ at which
\begin{equation}\label{eq:regularpoint}
    N_o(E +\delta) - N_o(E-\delta) \ \le \ \Gamma(E) \, \delta\; ,
\end{equation}
for some $\Gamma(E) < \infty$. In particular if $\eps$ is $C^1$ and $\nabla
\eps$ is non-zero on the level set $\set{\eps(\mathbf q) = E}$, then $E$ is a
regular point.  For example, with $H_o$ the discrete Laplacian on $\ell^2(\Z)$,
\begin{equation*}
  H_o \psi(x) \ = \ \psi(x + 1) + \psi(x-1) \; ,
\end{equation*}
we have the symbol $\eps(q) = 2 \cos ( q)$ and every $E \in (-2,2)$ is a
regular point. However at the band edges, $E = \pm 2$, the difference on the
right hand side of \eq{eq:regularpoint} is only $\mathcal O (\delta^{\half})$,
and these points are not regular points. We consider the behavior of
$N_\lambda(E)$ at such ``points of order $\alpha$,'' here $\alpha =1/2$, in
Theorem \ref{thm:orderalpha} below.

Our main result involves the density of states of $H_\lambda$ at a regular
point:
\begin{thm}\label{thm}
  Suppose $\int |\omega|^q \rho(\omega) \di \omega < \infty$
  for some $2 < q < \infty$ or that $\rho$ is compactly supported,
  in which case set $q=\infty$. If
  $E$ is a regular point for $\eps$, then there
  is $C_q = C_q(\rho, \Gamma(E)) < \infty$
  such that
  \begin{equation}\label{eq:mainbound}
    N_\lambda(E+\delta) - N_\lambda(E-\delta) \ \le \
    \Gamma(E) \, \delta \ + \ C_{q} \,
    \lambda^{\frac{1}{3} (1 + \frac{2}{q})}
    \delta^{\frac{1}{3}(1-\frac{2}{q})}
  \end{equation}
  for all $\lambda,\delta \ge 0$.
\end{thm}

For very small $\delta$, namely
\begin{equation*}
    \frac{\delta}{\lambda} \ \lesssim \ \lambda^{\frac{1}{3} (1 + \frac{2}{q})}
    \delta^{\frac{1}{3}(1-\frac{2}{q})} \; ,
\end{equation*}
the Wegner bound \eq{eq:Wegner} is stronger than \eq{eq:mainbound}.\footnote{We
  thank M. Disertori for this observation.}  Thus Theorem \ref{thm}
is useful only for
\begin{equation*}
    \delta \ \gtrsim \ \lambda^{\frac{2q+1}{q+1}} \; .
\end{equation*}
Combining the Wegner estimate and Theorem \ref{thm} for these separate regions
yields the following:
\begin{cor}\label{cor}
  Under the hypotheses of Theorem \ref{thm}, there is $C_q < \infty$, with
  $C_q = C_q( \rho,\Gamma(E))$, such that
  \begin{equation}
    N_\lambda (E+\delta)-N_\lambda(E-\delta) \ \le \ C_q
    \delta^{\half ( 1 - \frac{1}{2q+1} )}
  \end{equation}
  for all $\lambda,\delta \ge 0$.
\end{cor}
\noindent Thus, the integrated density of states is H\"older equi-continuous of
order $\half$ as $\lambda \rightarrow 0$ (if $\rho$ is compactly supported).

The starting point for our analysis of the density of states is a well known
formula relating $\di N_\lambda$ to the resolvent of $H_\omega$,
\begin{equation*}
    \frac{\di N_\lambda(E)}{\di E }\ = \ \lim_{\eta \downarrow 0} \frac{1}{\pi}\int_\Omega
    \Im \ip{\delta_0}{(H_\omega - E - \im \eta)^{-1} \delta_0} \di
    \Pr(\omega) \; .
\end{equation*}
The general idea of the proof is to express $\Im \ip{\delta_0}{(H_\omega - E -
\im \eta)^{-1} \delta_0}$ using a finite resolvent expansion to second order
\begin{multline}\label{eq:secondorder}
(H_\omega - E - \im \eta)^{-1} \\
\begin{aligned}
     &= \ (H_o - E - \im \eta)^{-1}
    - \lambda (H_o - E - \im \eta)^{-1} V_\omega (H_o - E - \im
    \eta)^{-1}\\
    &\quad + \  \lambda^2 (H_o - E - \im \eta)^{-1}
    V_\omega(H_\omega - E - \im \eta)^{-1} V_\omega
     (H_o - E - \im \eta)^{-1} \; ,
\end{aligned}
\end{multline}
and to use the Wegner bound \eq{eq:Wegner} to estimate the last term, with the
resulting factor of $1/\lambda$ controlled by the factor $\lambda^2$.

Here is a simplified version of the argument which works if $E$
falls outside the spectrum of $H_o$ and $\psi_E =(H_o
-E)^{-1}\delta_0 \in \ell^1(\Z^d)$. The first two terms of
\eq{eq:secondorder} are bounded and self-adjoint when $\eta = 0$,
so
\begin{multline*}
    \lim_{\eta \downarrow 0} \frac{1}{\pi}\int_\Omega
    \Im \ip{\delta_0}{(H_\omega - E - \im \eta)^{-1} \delta_0} \di
    \Pr(\omega) \\
\begin{aligned}
    &= \ \lambda^2 \lim_{\eta \downarrow 0} \frac{1}{\pi}\int_\Omega
    \Im \ip{ \psi_E }{V_\omega (H_\omega - E - \im \eta)^{-1} V_\omega \psi_E} \di
    \Pr(\omega) \\
    &\le \ \lambda^2 \lim_{\eta \downarrow 0} \sum_{x,y}
     \abs{\psi_E(x)}\abs{ \psi_E(y)}
\end{aligned}
\\ \times \frac{\eta}{\pi} \int_\Omega
    \abs{ \omega(x) \omega(y) \ip{\delta_x}{ \left ((H_\omega - E)^2 + \eta^2 \right )^{-1} \delta_y}} \di
    \Pr(\omega)  \; .
\end{multline*}
If $\rho$ is, say, compactly supported, then
\begin{multline*}
   \lim_{\eta \downarrow 0} \frac{\eta}{\pi} \int_\Omega
   \abs{ \omega(x) \omega(y) \ip{\delta_x}{ \left ((H_\omega - E)^2 + \eta^2 \right )^{-1}  \delta_y}}\di
    \Pr(\omega) \\
    \lesssim \ \lim_{\eta \downarrow 0} \frac{\eta}{\pi} \int_\Omega
   \ip{\delta_x}{ \left ((H_\omega - E)^2 + \eta^2 \right )^{-1}  \delta_y}\di
    \Pr(\omega) \
    \lesssim \ \frac{1}{\lambda} \; ,
\end{multline*}
by the Wegner bound, and therefore
\begin{equation}\label{eq:outsidespectrum}
    \frac{\di N_\lambda(E) }{\di E} \ \lesssim \ \lambda
    \norm{\psi_E}_1^2 \; , \quad \text{ for } E \not \in
    \sigma(H_o) \; .
\end{equation}
We have used second order perturbation theory to ``boot-strap'' the Wegner
estimate and obtain an estimate of lower order in $\lambda$. Unfortunately, as
$\rho$ was assumed compactly supported, $E$ is not in the spectrum of
$H_\lambda$ for sufficiently small $\lambda$, and thus $\di N_\lambda(E)/\di E
= 0$. So, in practice, \eq{eq:outsidespectrum} is not a useful bound.

Nonetheless, in the cases covered by Theorem \ref{thm}, $H_\lambda$ can have
spectrum in a neighborhood of $E$, even for small $\lambda$, since $E$ may be
in the interior of the spectrum of $H_o$.  Although, the above argument does
not go through, we shall exploit the translation invariance of the distribution
of $H_\omega$ by introducing a Fourier transform on the Hilbert space of
``random wave functions,'' complex valued functions $\Psi(x,\omega)$ of
$(x,\omega) \in \ell^2(\Z^d) \times \Omega$ with
\begin{equation*}
    \sum_x \int_\Omega \abs{\Psi(x,\omega)}^2 \di \Pr(\omega) <
    \infty \; .
\end{equation*}
Under this Fourier transform an integral $\int_\Omega$ of a matrix element of
$f(H_\omega)$ is replaced by an integral $\int_{T^d}$ over the $d$-torus of a
matrix element of $f(\widehat H_{\mathbf k})$, with $\widehat H_{\mathbf k}$ a
certain family of operators on $L^2(\Omega)$ (see \eq{eq:magicformula}). Off
the set $S_\eps := \{ \mathbf k \in T^d | |\eps(\mathbf k)-E|
> \epsilon \}$ with $\epsilon
>> \delta$, we are able to carry out an argument similar to that which
led to \eq{eq:outsidespectrum}.  To prove Theorem \ref{thm}, we
shall directly estimate
\begin{equation*}
    N(E+\delta) - N(E - \delta) \ = \ \int_\Omega
    \ip{\delta_0}{P_\delta(H_\omega) \delta_0} \di \Pr (\omega) \;
    ,
\end{equation*}
with $P_\delta$ the characteristic function of the interval $[E-\delta, E +
\delta]$, because the integrand on the r.h.s. is bounded by $1$.  Since $E$ is
a regular point, the error in restricting to $S_\eps$ will be bounded by
$\Gamma(E) \eps$. Choosing $\eps$ optimally will lead to Theorem \ref{thm}.

More generally, we say that $E$ is a point of order $\alpha$ for
$\eps$, if there exists $\Gamma(E;\alpha)$ such that
\begin{equation*}
    N_o(E + \delta) - N_o(E-\delta) \ \le \ \Gamma(E;\alpha)
    \delta^\alpha \; .
\end{equation*}
If $E \not \in \sigma(H_o)$, we say that $E$ is a point of order $\infty$ and
set $\Gamma(E;\infty) =0$. For points of order $\alpha$ we have the following
extension of Theorem \ref{thm}.
\begin{thm}\label{thm:orderalpha}
  Suppose $\int |\omega|^q \rho(\omega) \di \omega < \infty$
  for some $2 < q < \infty$ or that $\rho$ is compactly supported,
  in which case set $q=\infty$. If
  $E$ is a point of order $\alpha \le \infty$ for $\eps$, then there
  is $C_{q,\alpha} = C_{q,\alpha}(\rho, \Gamma(E;\alpha)) < \infty$
  such that
  \begin{equation}\label{eq:mainbound3}
    N_\lambda(E+\delta) - N_\lambda(E-\delta) \ \le \ \Gamma(E;\alpha)
    \delta^\alpha \ + \
    C_{q,\alpha} \,
    \left [ \lambda^{1 + \frac{2}{q} }
    \delta^{1-\frac{2}{q} } \right ]^\frac{1}{1+ \frac{2}{\alpha}}
  \end{equation}
  for all $\lambda,\delta \ge 0$.
\end{thm}

When $\alpha = \infty$ and $q=\infty$, so $E \not \in \sigma(H_o)$ and $\rho$
is compactly supported, the result is technically true but uninteresting since
$E \not \in \sigma(H_\lambda)$ for small $\lambda$, as discussed above. However
for $q < \infty$, we need not have that $\rho$ is compactly supported, and
$E\not \in \sigma(H_o)$ may still be in the spectrum of $H_\lambda$ for
arbitrarily small $\lambda$. In this case, \eq{eq:mainbound3} signifies that
\begin{equation*}
    N_\lambda(E+\delta) - N_\lambda(E-\delta) \ \le \
    C_{q,\infty}  \lambda^{1 + \frac{2}{q} }
    \delta^{1-\frac{2}{q} } \; ,
\end{equation*}
which in fact improves on the Wegner bound for appropriate $\lambda, \delta$.

As above, we may use the Wegner bound for $\delta$ very small to improve on
\eq{eq:mainbound3}:
\begin{cor}
  Under the hypotheses of Theorem \ref{thm:orderalpha}, there is $C_{q,\alpha} =
  C_{q,\alpha}(\rho,\Gamma(E;\alpha)) < \infty $ such that
  \begin{equation*}
    N_\lambda (E+\delta)-N_\lambda(E-\delta) \ \le \ C_{q;\alpha}
    \delta^{\frac{\alpha}{\alpha + 1} \left ( 1 - \frac{1}{\frac{\alpha + 1}{\alpha}q+1} \right )}
  \end{equation*}
  for all $\lambda,\delta \ge 0$.
\end{cor}

The inspiration for these results is the (non-rigorous)
renormalized perturbation theory for $\di N_\lambda$ which has
appeared in the physics literature, e.g., ref.~\cite{Thouless1984}
and references therein. If $\int \omega \rho(\omega) \di \omega
=0$ and $\int \omega^2 \rho(\omega) \di \omega = 1$, as can always
be achieved by shifting the origin of energy and re-scaling
$\lambda$, then the central result of that analysis is that
\begin{equation*}
    \frac{\di N_\lambda(E)}{\di E } \ \approx \ \frac{1}{\pi} \Im
    \ip{\delta_0}{\left ( H_o - E - \lambda^2
    \Gamma_\lambda(E) \right )^{-1}\delta_0} \; ,
\end{equation*}
where $\Gamma_\lambda(E)$, the so-called ``self energy,'' satisfies $\Im \,
\Gamma_\lambda(E) > 0$ with
\begin{equation*}
    \lim_{\lambda \rightarrow 0} \Im \, \Gamma_\lambda(E) \ \approx \
    \lim_{\eta \rightarrow 0}
    \Im \ip{\delta_0}{\left ( H_o - E - \im \eta \right )^{-1} \delta_0}
    \ = \ \pi \frac{\di N_0(E)}{\di E } \; .
\end{equation*}

Up to a point, the self-energy analysis may be followed
rigorously.  Specifically, one can show (see \S\ref{sec:Fourier}):
\begin{prop}\label{prop}
If $\int \omega \rho(\omega) \di \omega = 0$ and $\int \omega^2
\rho(\omega) \di \omega =1$, then for each $\lambda
>0$ there is a map $\Gamma_\lambda$ from  $\{ \Im z > 0 \}$ to the translation invariant operators with
non-negative imaginary part on $\ell^2(\Z^2)$ such that
\begin{equation}\label{eq:selfenergy}
    \int_\Omega
    (H_\omega - z)^{-1}  \di
    \Pr(\omega) \ = \ \left (H_o - z -
    \lambda^2 \Gamma_\lambda(z) \right )^{-1}
    \; ,
\end{equation}
and for {\em fixed} $z\in \{ \Im z > 0\}$
\begin{equation}\label{eq:limitofselfenergy}
    \lim_{\lambda \rightarrow 0} \ip{\delta_x}{\Gamma_\lambda(z)\delta_y}
    \ = \
    \ip{\delta_0}{(H_o-z)^{-1}
\delta_0} \, \delta_{x,y}\; .
\end{equation}
\end{prop}
\noindent However there is {\em a priori} no uniformity in $z$ for
the convergence in \eq{eq:limitofselfenergy}, so for fixed
$\lambda$ we may conclude nothing about
\begin{equation*}
    \lim_{\eta \downarrow 0} \left (H_o - E - \im \eta - \lambda^2
    \Gamma_\lambda(E + \im \eta) \right )^{-1} \; .
\end{equation*}

Still, one is left feeling that Theorem \ref{thm} and Corollary
\ref{cor} are not-optimal, and the ``standard wisdom'' is that
something like the following is true.
\begin{conj}
  Let $\rho$ have moments of all orders, i.e., $\int |\omega|^q \rho(\omega) < \infty$ for
  all $q \ge
  1$.  Given $E_o \in \R$, if there is $\delta > 0$ such that
  on the set $\{ \mathbf{q} :
  \abs{\eps(\mathbf q) - E_o} < \delta\}$ the symbol $\eps$ is
  $C^1$ with $\nabla \eps(\mathbf q) \neq 0$, then there is $C_\delta  < \infty$ such that
  \begin{equation*}
    \frac{\di N_\lambda(E)}{\di E} \ \le \ C_\delta
  \end{equation*}
  for all $\lambda \in \R$ and $E \in [E_o-\half \delta, E_o + \half \delta]$.
\end{conj}
\noindent {\bf Remark:} The requirement that $\rho$ have moments of all orders
is simply the minimal requirement for the infinite perturbation series for
$(H_o-z - \lambda V_\omega)^{-1}$ to have finite expectation at each order (for
$\Im z > 0$).  In fact, this may be superfluous, as suggested by the example of
Cauchy randomness, for which the density of states can be explicitly computed,
see ref.~\cite{Spencer1984}:
\begin{equation*}
    \di N_\lambda(E) \ = \ \frac{1}{\pi} \int_{T^d}
    \frac{\lambda}{(\eps(\mathbf q) - E)^2 + \lambda^2} \diq
     \; , \quad \text{for } \rho(\omega) = \frac{1}{\pi}
    \frac{1}{1 + \omega^2}\; ,
\end{equation*}
although $\int \rho(\omega) \abs{\omega}^q = \infty$ for every $q \ge 1$.

\section{Translation invariance, augmented space, and a Fourier transform}
\label{sec:Fourier} The joint probability measure $\Pr(\omega)$
for the random function $\omega:\Z^d \rightarrow \R$ is
\begin{equation*}
    \di \Pr(\omega) \ := \ \prod_{x \in \Z^d} \rho(\omega(x)) \di \omega(x)
\end{equation*}
on the probability space $\Omega =  \R^{\Z^d}$. Clearly, $\Pr(\omega)$ is
invariant under the translations $\tau_\xi : \Omega \rightarrow \Omega$ defined
by
\begin{equation*}
    \tau_\xi \omega(x) \ = \ \omega(x - \xi) \; .
\end{equation*}
In particular, since
\begin{equation}\label{eq:distinv}
    S_\xi H_\omega S_\xi^\dag \ = \ H_o + V_{\tau_{\xi} \omega} \
    = \ H_{\tau_{\xi} \omega} \; ,
\end{equation}
$H_\omega$ and $S_\xi H_\omega S_\xi^\dag$ are identically
distributed for any $\xi \in \Z^d$,

To express this invariance in operator theoretic terms, we
introduce the fibred action of $H_\omega$ on the Hilbert space
$L^2(\Omega;\ell^2(\Z^d))$ -- the space of ``random wave
functions'' -- namely,
\begin{equation*}
    \Psi(\omega) \ \mapsto \ H_\omega \Psi(\omega) \; .
\end{equation*}
We identify $L^2(\Omega;\ell^2(\Z^d))$ with $L^2(\Omega \times \Z^d)$ and
denote the action of $H_\omega$ on the latter space by $\mathbf H$, so
\begin{equation*}
    [{\mathbf H} \Psi](\omega,x) \ = \ \sum_{\xi} \check{\eps}(\xi)
    \Psi(\omega,x-\xi) \ + \ \lambda \omega(x) \Psi(\omega,x) \;
    .
\end{equation*}
The following elementary identity relates $\int_\Omega f(H_\omega) \di
\Pr(\omega)$ to $f(\mathbf H)$, for any bounded measurable function $f$,
\begin{equation}\label{eq:basicformula}
    \int_\Omega \di \Pr(\omega) \left < \delta_x, f(H_\omega) \delta_y\right > \ = \ \left <
    \E^\dag \delta_x,  f({\mathbf H}) \E^\dag \delta_y\right > \; ,
\end{equation}
where $\E^\dag$ is the adjoint of the linear expectation map $\E :
L^2(\Omega \times \Z^d) \rightarrow \ell^2(\Z^d)$ defined by
\begin{equation*}
    [\E \Psi](x) \ = \ \int_\Omega \Psi(\omega,x) \di \Pr(\omega)
    \; .
\end{equation*}
Note that $\E^\dag$ is an isometry from $\ell^2(\Z^d)$ onto the subspace of
functions independent of $\omega$ -- ``non-random functions."

The general fact that averages of certain quantities depending on $H_\omega$
can be represented as matrix elements of $\mathbf H$ is known, and is sometimes
called the ``augmented space representation'' (e.g., ref.
\cite{Mookerjee1973_1,Mookerjee1973_2,KaplanGray1977}) where ``augmented
space'' refers to the Hilbert space $L^2(\Omega \times \Z^d)$. There are
``augmented space'' formulae other than \eq{eq:basicformula}, such as
\begin{equation}\label{eq:basicformula2}
    \int_\Omega \di \Pr(\omega) \omega(x) \omega(y)
    \left < \delta_x, f(H_\omega) \delta_y\right > \ = \
    \left <
    \E^\dag \delta_x,  {\mathbf V} f({\mathbf H}) {\mathbf V} \E^\dag \delta_y\right
    > \; ,
\end{equation}
and
\begin{equation*}
    \int_\Omega \di \Pr(\omega) \left < \delta_x, f(H_\omega) \delta_0\right >
    \left < \delta_0, g(H_\omega) \delta_y\right > \ = \
    \left <
    \E^\dag \delta_x,  f({\mathbf H}) P_0 g(\mathbf H) \E^\dag \delta_y\right
    >\; ,
\end{equation*}
where $P_0$ denotes the projection $P_0\Psi(\omega,x) \ = \ \Psi(\omega,0)$ if
$x=0$ and $0$ otherwise.  The first of these (\eq{eq:basicformula2}) will play
a roll in the proof of Theorem \ref{thm}.

There are two natural groups of unitary translations on $L^2(\Omega \times
\Z^d)$:
\begin{equation*}
    S_\xi \Psi(\omega,x) = \Psi(\omega,x-\xi) \; ,
\end{equation*}
and
\begin{equation*}
      T_\xi \Psi(\omega,x) \ = \ \Psi(\tau_{-\xi}\omega,x) \; .
\end{equation*}
Note that these groups commute: $\com{S_\xi}{T_{\xi'}} = 0$ for every $\xi,\xi'
\in \Z^d$. A key observation is that the distributional invariance of
$H_\omega$, \eq{eq:distinv}, results in the {\em invariance} of ${\mathbf H}$
under the combined translations $T_\xi S_\xi = S_\xi T_\xi$:
\begin{equation*}
    S_\xi T_\xi {\mathbf H} T_\xi^\dag S_\xi^\dag  \ = \ {\mathbf
    H}\; .
\end{equation*}
In fact, let us define
\begin{equation*}
    \mathbf H_o \ = \ \sum_{\xi} \check{\eps}(\xi) S_\xi \; ,
    \quad \mathbf V \Psi(\omega,x) \ = \ \omega(x) \Psi(\omega,x)
    \; .
\end{equation*}
Then ${\mathbf H} = \mathbf H_o + \lambda \mathbf V$ where $\mathbf H_o$
commutes with $S_\xi$ and $T_\xi$ while for $\mathbf V$ we have
\begin{equation*}
     \mathbf V S_\xi \ = \ T_{-\xi}\mathbf V \; .
\end{equation*}

To exploit this translation invariance of ${\mathbf H}$, we define a Fourier
transform which diagonalizes the translations $S_\xi T_\xi$ (and therefore
partially diagonalizes ${\mathbf H}$). The result is a unitary map ${\mathcal
F}: L^2(\Omega \times \Z^d) \rightarrow L^2(\Omega \times T^d)$, with $T^d$ the
$d$-torus $[0,2 \pi)^d$. Let us define $\mathcal F$ first on functions having
finite support in $\Z^d$ by
\begin{equation*}
    \mathcal F \Psi (\omega,\mathbf k) \ = \ \sum_{\xi} \e^{-\im \mathbf k \cdot \xi}
    \Psi(-\xi,\tau_{-\xi}\omega) \; .
\end{equation*}
It is easy to verify, using well known properties of the usual Fourier series
mapping $\ell^2(\Z^d) \rightarrow L^2(T^d)$, that $\mathcal F$ extends to a
unitary isomorphism $L^2(\Omega \times \Z^d) \rightarrow L^2(\Omega \times
T^d)$, i.e. that $\mathcal F \mathcal F^\dag =\1$ and $\mathcal F^\dag \mathcal
F = \1$ where $\mathcal F ^\dag$ is the adjoint map
\begin{equation*}
    \mathcal F^\dag\widehat \Psi(\omega,x) \ = \ \int_{T^d}
    \e^{-\im \mathbf k \cdot x} \widehat \Psi(\tau_{-x} \omega, \mathbf
    k)\dik \; .
\end{equation*}

Another way of looking at $\mathcal F$ is to define for each
${\mathbf k} \in T^d$ an operator $\mathcal F_{\mathbf k}:
L^2(\Omega \times \Z^d) \rightarrow L^2(\Omega)$ by
\begin{equation*}
    \mathcal F_{\mathbf k} \Psi \ =  \ \lim_{L \rightarrow \infty}
    \sum_{|\xi| < L} \e^{-\im \mathbf k \cdot
    \xi} {\mathcal J} S_\xi T_\xi \Psi \; ,
\end{equation*}
where ${\mathcal J}$ is the evaluation map $\mathcal J \Psi(\omega) \ = \
\Psi(\omega,0)$. The maps $\mathcal F_{\mathbf k}$ are {\em not} bounded, but
are densely defined with $\mathcal F_{\mathbf k}
    \Psi \in L^2(\Omega)$ for almost every $\mathbf k$, and
\begin{equation*}
    \mathcal F \Psi(\omega, \mathbf k) \ = \ \mathcal F_{\mathbf k}
    \Psi(\omega) \quad \mbox{a.e. $\omega$, $\mathbf k$}.
\end{equation*}
If we look at $L^2(\Omega \times T^d)$ as the direct integral $\int^\oplus \din
\mathbf k L^2(\Omega) $, then
\begin{equation*}
    \mathcal F = \int^\oplus \mathcal \din \mathbf k  \mathcal F_{\mathbf k} \; .
\end{equation*}

This Fourier transform diagonalizes the combined translation
$S_\xi T_\xi$,
\begin{equation*}
    \mathcal F_{\mathbf k} S_\xi T_\xi  \ = \ \e^{\im
    \mathbf k \cdot \xi} \mathcal F_{\mathbf k} \; ,
\end{equation*}
as follows from the following identities for $S$ and $T$,
\begin{equation*}
    \mathcal F_{\mathbf k} T_\xi  \ =  \ T_\xi \mathcal F_{\mathbf k} \; ,  \quad
    \mathcal F_{\mathbf k} S_\xi  \ =  \  \e^{\im
    \mathbf k \cdot \xi} T_{-\xi} \mathcal F_{\mathbf k} \; ,
\end{equation*}
where, on the right hand sides, $T_\xi$ denotes the operator $
    T_\xi\psi(\omega)  =  \psi(\tau_{-\xi}\omega)$
on $L^2(\Omega)$. Furthermore, explicit computation shows that
\begin{equation*}
    \mathcal F_{\mathbf k} \mathbf V \ = \ \omega(0) \mathcal F_{\mathbf
    k}\; ,
\end{equation*}
where $\omega(0)$ denotes the operator of multiplication by the random variable
$\omega(0)$, $\psi(\omega) \ \mapsto \ \omega(0) \psi(\omega)$.  Putting this
all together yields
\begin{prop}
  Under the natural identification of $L^2(\Omega,T^d)$ with the direct integral
  $\int^\oplus \din \mathbf k L^2(\Omega)$, the operator
  $\mathbf {\widehat H} = \mathcal F \mathbf H \mathcal
  F^\dag$ is partially diagonalized, $
  \mathbf {\widehat H} = \int^\oplus \widehat H_{\mathbf k}$,
  with $\widehat H_{\mathbf k}$ operators on
  $L^2(\Omega)$ given by the following formula
  \begin{equation*}
    \widehat H_{\mathbf k}  \ = \ \sum_{\xi}
    \e^{-\im \mathbf k \cdot \xi} \check \eps(-\xi)
    T_\xi \ + \ \lambda \omega(0) \; .
  \end{equation*}
\end{prop}
Let us introduce for each $\mathbf k \in T^d$,
\begin{equation*}
\widehat H_{\mathbf k}^o \ := \ \sum_{\xi}
    \e^{-\im \mathbf k \cdot \xi} \check \eps(-\xi)
    T_\xi \ = \ \sum_{\xi} \left [  \int_{T^d} \eps(\mathbf q+\mathbf k)
    \e^{\im \xi \cdot \mathbf q}
     \, \diq \right ] T_\xi \; ,
\end{equation*}
so $\widehat H_{\mathbf k} = \widehat H_{\mathbf k}^o + \lambda \omega(0)$.
Note that
\begin{equation*}
    \widehat H_{\mathbf k}^o \chi_\Omega \ = \ \eps(\mathbf k) \chi_\Omega
    \; ,
\end{equation*}
where $\chi_\Omega(\omega)=1$ for every $\omega \in \Omega$.  That is,
$\chi_\Omega$ is an eigenvector for $H_{\mathbf k}^o$.\footnote{In fact, if
$\eps$ is almost everywhere non-constant (so $H_o$ has no eigenvalues) then
$\eps(k)$ is the {\em unique} eigenvalue for $\widehat H_{\mathbf k}^o$ and the
remaining spectrum of $\widehat H_{\mathbf k}^o$ is infinitely degenerate
absolutely continuous spectrum. One way to see this is to let $\phi_n(v)$ be
the orthonormal polynomials with respect the weight $\rho(v)$, and look at the
action of $\widehat H_{\mathbf k}^o$ on the basis for $L^2(\Omega)$ consisting
of products of the form $\prod_{x \in \Z^d} \phi_{n(x)}(\omega(x))$ with only
finitely many $n(x) \neq 0$.}

Applying the Fourier transform $\mathcal F$ to the right hand side
of the ``augmented space'' formula \eq{eq:basicformula} we obtain
the following beautiful identity, central to this work:
\begin{equation}\label{eq:magicformula}
  \int_\Omega \di \Pr(\omega) \ip{\delta_x}{f(H_\omega) \delta_y}
  \ = \ \int_{T^d} \dik \e^{\im {\mathbf k} \cdot (x-y)}\ip{\chi_\Omega}
    {f(\widehat H_{\mathbf k})
  \chi_\Omega} \; .
\end{equation}
Similarly, we obtain
\begin{multline}\label{eq:magicformula2}
    \int_\Omega \di \Pr(\omega) \omega(x) \omega(y) \ip{\delta_x}{f(H_\omega) \delta_y}
    \\ = \ \int_{T^d} \dik \e^{\im {\mathbf k} \cdot (x-y)}\ip{\omega(0)\chi_\Omega}{f(\widehat H_{\mathbf k})
  \omega(0)\chi_\Omega} \;
\end{multline}
from \eq{eq:basicformula2}. Related formulae have been used, for example, to
derive the Aubry duality between strong and weak disorder for the almost
Mathieu equation, see ref.~\cite{GordonJitomirskayaLastSimon1997} and
references therein.

As a first application of \eq{eq:magicformula}, let us prove the
existence of the self energy (Prop. \ref{prop}) starting from the
identity
\begin{equation*}
    \int_\Omega \di \Pr(\omega) \ip{\delta_0}{\left ( H_\omega - z \right )^{-1} \delta_0}
    \ = \ \int_{T^d} \dik \ip{\chi_\Omega}{ \left (
    \widehat H_{\mathbf k} - z \right )^{-1}\chi_\Omega} \; .
\end{equation*}

\begin{proof}[Proof of Prop. \ref{prop}]
Since $\chi_\Omega$ is an eigenvector of $\widehat H_{\mathbf
k}^o$ and
\begin{equation*}
\ip{\chi_\Omega}{\omega(0) \chi_\Omega} \ = \ \int \omega \rho(\omega) \di
\omega \ = \ 0 \; ,
\end{equation*}
the Feschbach mapping implies
\begin{equation}\label{eq:selfenergy2}
    \ip{\chi_\Omega}{ \left (
    \widehat H_{\mathbf k} - z \right )^{-1}\chi_\Omega} \ = \
    \left ( \eps(k)  - z -
    \lambda^2 \Gamma_\lambda(z;\mathbf k)
    \right )^{-1} \; ,
\end{equation}
with
\begin{equation*}
    \Gamma_\lambda(z;\mathbf k) \ = \ \ip{\omega(0)\chi_\Omega}{\left (P^\perp \widehat
    H_{\mathbf k} P^\perp - z \right )^{-1} \omega(0) \chi_\Omega}
    \; ,
\end{equation*}
where $P^\perp$ denotes the projection onto the orthogonal complement of
$\chi_\Omega$ in $L^2(\Omega)$.

Let the self energy $\Gamma_\lambda(z)$ be the translation
invariant operator with symbol $\Gamma_\lambda(z;\mathbf k)$,
i.e.,
\begin{equation*}
    \ip{\delta_x}{\Gamma_\lambda(z)\delta_y} \ = \ \int_{T^d}
    \e^{\im \mathbf k \cdot (x-y)} \Gamma_\lambda(z;\mathbf k)
    \dik \; .
\end{equation*}
Clearly $\Gamma_\lambda(z)$ is bounded with non-negative imaginary part.
Furthermore by \eq{eq:magicformula} and \eq{eq:selfenergy2}, the identity
\eq{eq:selfenergy} holds, namely
\begin{equation*}
\int_\Omega
    (H_\omega - z)^{-1}  \di
    \Pr(\omega) \ = \ \left (H_o - z - \lambda^2 \Gamma_\lambda(z)
    \right )^{-1} \; .
\end{equation*}
It is clear that
\begin{equation*}
    \lim_{\lambda \rightarrow 0}
    \Gamma_\lambda(z;\mathbf k) \ = \ \ip{\omega(0)\chi_\Omega}{\left (\widehat
    H_{\mathbf k}^o - z \right )^{-1} \omega(0) \chi_\Omega} \; ,
\end{equation*}
from which \eq{eq:limitofselfenergy} follows easily.
\end{proof}

\section{Proofs}
We first prove Theorem \ref{thm} and then describe modifications of the proof
which imply Theorem \ref{thm:orderalpha}.
\subsection{Proof of Theorem \ref{thm}}
Fix a regular point $E$ for $\eps$, and for each $\delta
>0$ let
\begin{multline*}
  f_\delta(t) \ = \ \half \left ( \chi_{(E-\delta,E+\delta)}(t) +
\chi_{[E-\delta, E+\delta]}(t) \right ) \\ = \ \begin{cases}   1 \, , & t\,
, \in (E-\delta, E+\delta) \, ,\\
\half & t=E\pm \delta \, ,\\
0 & t \not \in [E-\delta,E+\delta] \, .
\end{cases}
\end{multline*}
Since $N_\lambda(E)$ is continuous (see \eq{eq:Lipschitz}),
\begin{equation*}
    N_\lambda(E + \delta) - N_\lambda(E - \delta) \ = \
    \int_\Omega \ip{\delta_0}{f_\delta(H_\omega) \delta_0} \di
    \Pr(\omega) \; .
\end{equation*}
Thus, in light of \eq{eq:magicformula}, our task is to show that
\begin{equation}\label{eq:toprove}
     \int_{T^d} \ip{\chi_\Omega}{f_\delta(\widehat H_{\mathbf k}) \chi_\Omega}
     \dik \ \le \
    \Gamma(E) \, \delta \ + \ C_{q} \,
    \lambda^{\frac{1}{3} (1 + \frac{2}{q})}
    \delta^{\frac{1}{3}(1-\frac{2}{q})} \; ,
\end{equation}
with a constant $C_q$ independent of $\delta$ and $\lambda$.  Note
that for each $\mathbf k \in T^d$
  \begin{equation*}
    \abs{\ip{\chi_\Omega}{f_\delta(\widehat H_{\mathbf k}) \chi_\Omega} }
    \ \le \ 1 \; ,
  \end{equation*}
so we can afford to neglect a set of Lebesgue measure $\lambda^{\frac{1}{3} (1
+ \frac{2}{q})} \delta^{\frac{1}{3}(1-\frac{2}{q})}$ on the l.h.s. of
\eq{eq:toprove}.

Consider $\mathbf k \in T^d$ with $\abs{\eps(\mathbf k) -E} > \delta$.  Then
$$f_\delta(\widehat H_{\mathbf k}^o) \chi_\Omega \ = \ f_\delta(\eps(\mathbf k))
\chi_\Omega \ = \ 0 \; .$$Thus
\begin{gather}\label{eq:step1}
\begin{split}
    & \ip{\chi_\Omega}{f_\delta(\widehat H_{\mathbf k}) \chi_\Omega} \ = \
    \ip{\chi_\Omega}{\left ( f_\delta(\widehat H_{\mathbf k}) -f_\delta(\widehat H^o_{\mathbf k}) \right )
    \chi_\Omega} \\
    & =  \ \lim_{\eta \rightarrow 0}  \frac{1}{\pi} \int_{E-
    \delta}^{E+\delta}  \Im  \ip{\chi_\Omega}{\left ( \frac{1}{\widehat H_{\mathbf
    k}-t - \im \eta} - \frac{1}{\widehat H^o_{\mathbf k} - t - \im
    \eta} \right ) \chi_\Omega} \di t  \\
    & = \ \lambda\lim_{\eta \rightarrow 0}  \frac{1}{\pi}\int_{E-
    \delta}^{E+\delta}   \Im \frac{1}{  t +\im \eta-\eps(\mathbf k)}
    \ip{\chi_\Omega}{\frac{1}{\widehat H_{\mathbf
    k}-t - \im \eta} \omega(0) \chi_\Omega} \di t \\
    & = \  \lambda
    \ip{\chi_\Omega}{\frac{1}{\widehat H_{\mathbf
    k} - \eps(\mathbf k) } f_\delta(\widehat H_{\mathbf
    k}) \omega(0) \chi_\Omega} \; ,
\end{split}
\intertext{since $(t- \eps(\mathbf k) )^{-1}$ is continuous for $t \in
[E-\delta, E+\delta]$. Using again that $f_\delta(\widehat H^o_{\mathbf k})
\chi_\Omega =0$, we find that the final term of \eq{eq:step1} equals}
\label{eq:step2}
\begin{split}
  & = \ip{\left [ \frac{1}{\widehat H_{\mathbf
    k}-\eps(\mathbf k)  } f_\delta(\widehat H_{\mathbf
    k}) - \frac{1}{ \widehat H^o_{\mathbf
    k}-\eps(\mathbf k)} f_\delta(\widehat H^o_{\mathbf
    k}) \right ] \chi_\Omega}{ \omega(0) \chi_\Omega} \\
      & = \lambda \lim_{\eta \rightarrow 0} \frac{1}{\pi} \int_{E-
    \delta}^{E+\delta} \frac{1}{t-\eps(\mathbf k)}
     \Im \frac{1}{  t +\im \eta-\eps(\mathbf k)}
    \\ & \qquad  \qquad \qquad \qquad \qquad \qquad
    \times \ip{\frac{1}{\widehat H_{\mathbf
    k}-t - \im \eta} \omega(0)\chi_\Omega}{ \omega(0) \chi_\Omega} \di t \\
    & = \ \lambda \ip{\omega(0)\chi_\Omega}{\frac{f_\delta(\widehat H_{\mathbf
    k})}{(\widehat H_{\mathbf
    k}-\eps(\mathbf k) )^2}  \omega(0) \chi_\Omega} \; .
    \end{split}
\end{gather}
Putting eqs.~\eqref{eq:step1} and \eqref{eq:step2} together yields
\begin{multline*}
    \ip{\chi_\Omega}{f_\delta(\widehat H_{\mathbf k}) \chi_\Omega} \ = \
  \lambda^2 \ip{\omega(0)\chi_\Omega}{\frac{f_\delta(\widehat H_{\mathbf
    k})}{(\widehat H_{\mathbf
    k}-\eps(\mathbf k) )^2}  \omega(0) \chi_\Omega} \\
    \le \ \lambda^2 \frac{1}{(|\eps(\mathbf k)-E |-\delta)^2}
    \ip{\omega(0)\chi_\Omega}{f_\delta(\widehat H_{\mathbf k})
    \omega(0)\chi_\Omega}\; .
\end{multline*}

Thus, for any $\epsilon > \delta$,
\begin{multline*}
\int_{\set{|\eps(\mathbf k)-E|>\epsilon} }
    \ip{\chi_\Omega}{f_\delta(\widehat H_{\mathbf k}) \chi_\Omega} \\
\begin{aligned}
       & \le
    \ \lambda^2 \frac{1}{(\epsilon-\delta)^2} \int_{T^d}
    \ip{\omega(0)\chi_\Omega}{f_\delta(\widehat H_{\mathbf k})
    \omega(0)\chi_\Omega} \\
    & =  \ \lambda^2 \frac{1}{(\epsilon- \delta)^2} \int_{\Omega}
    \omega(0)^2 \ip{\delta_0}{f_\delta(H_{\omega})\delta_0} \di
    \Pr(\omega) \; ,
\end{aligned}
\end{multline*}
where in the last equality we have inverted the Fourier transform, using
\eq{eq:magicformula2}. We may estimate the right hand side with H\"older's
inequality and the Wegner estimate:
\begin{multline*}
    \int_\Omega\omega(0)^2\ip{\delta_0}{f_\delta(H_{\omega})\delta_0} \di
    \Pr(\omega) \\
    \begin{aligned} & \le  \
    \norm{\omega(0)}_q^2 \left ( \int_{\Omega}
    \ip{\delta_0}{f_{\delta}(H_{\omega})\delta_0} \di
    \Pr(\omega) \right )^{1- \frac{2}{q}} \\
    & \le  \ \norm{\omega(0)}_q^2
    \left ( \frac{\norm{\rho}_\infty}{\lambda}\, {2\delta} \right )^{1-
    \frac{2}{q}} \; , \end{aligned}
\end{multline*}
since $\ip{\delta_0}{f_{\delta}(H_{\omega})\delta_0}^p \le
\ip{\delta_0}{f_{\delta}(H_{\omega})\delta_0}$ for $p >1$ (because
$\ip{\delta_0}{f_{\delta}(H_{\omega})\delta_0}$  $\le 1$). Here
$\norm{\omega(0)}_q^q = \int \omega(0)^q \di \Pr(\omega)$ for $q  < \infty$ and
$\norm{\omega(0)}_\infty = \esssup_{\omega} \abs{\omega(0)}$.

Therefore
\begin{equation}\label{eq:finalresult}
    \int_{T^d}
    \ip{\chi_\Omega}{f_\delta(\widehat H_{\mathbf k}) \chi_\Omega}
    \ \le \ \Gamma(E) \epsilon  \ + \ \lambda^2
    \frac{1}{(\epsilon-\delta)^2} \norm{\omega(0)}_q^2
    \left ( \frac{\norm{\rho}_\infty}{\lambda}\, 2\delta \right )^{1-
    \frac{2}{q}}  \; ,
\end{equation}
where the first term on the right hand side is an upper bound for
\begin{equation*}
    \int_{\set{|\eps(\mathbf k)-E|\le \epsilon} }
    \ip{\chi_\Omega}{f_\delta(\widehat H_{\mathbf k}) \chi_\Omega}
    \dik
    \ \le \ \int_{\set{|\eps(\mathbf k)-E|\le \epsilon} } \dik \;
    .
\end{equation*}
Upon optimizing over $\epsilon \in ( \delta,\infty)$, this implies
\begin{equation*}
    \int_{\Omega}
    \ip{\delta_0 }{f_\delta( H_{\omega}) \delta_o} \ \le
    \ \Gamma(E) \delta \ + \ C_{\rho,q,\Gamma} \,
    \lambda^{\frac{1}{3} (1 + \frac{2}{q})}
    \delta^{\frac{1}{3}(1-\frac{2}{q})} \; ,
\end{equation*}
which completes the proof of Theorem \ref{thm}. \qed

\subsection{Proof of Theorem \ref{thm:orderalpha}}
If instead of being a regular point, $E$ is a point of order
$\alpha$ then the proof goes through up to \eq{eq:finalresult}, in
place of which we have
\begin{equation*}
    \int_{T^d}
    \ip{\chi_\Omega}{f_\delta(\widehat H_{\mathbf k}) \chi_\Omega}
    \ \le \ \Gamma(E;\alpha) \epsilon^\alpha   +  \lambda^2
    \frac{1}{(\epsilon-\delta)^2} \norm{\omega(0)}_q^2
    \left ( \frac{\norm{\rho}_\infty}{\lambda}\, \delta \right )^{1-
    \frac{2}{q}}  \; .
\end{equation*}
Setting $\eps = \delta + \lambda^\gamma \delta^\beta$ and choosing
$\gamma,\beta$ such that the two terms are of the same order yields
\begin{equation*}
    \gamma \ = \ \frac{1}{2 + \alpha} \left ( 1 + \frac{2}{q}
    \right ) \; , \qquad \beta \ = \ \frac{1}{2 + \alpha} \left ( 1 - \frac{2}{q}
    \right ) \; ,
\end{equation*}
which implies
\begin{equation*}
   \int_{\Omega}
    \ip{\delta_0 }{f_\delta( H_{\omega}) \delta_o} \ \le \ \Gamma(E;\alpha)
    \delta^\alpha \ +
    \ C_q \, \lambda^{\frac{\alpha}{2 + \alpha} \left ( 1 + \frac{2}{q}
    \right )} \ \delta^{\frac{\alpha}{2 + \alpha} \left ( 1 - \frac{2}{q}
    \right )} \; ,
\end{equation*}
completing the proof. \qed

\subsection*{Acknowledgements} I thank G.M. Graf and M. Disertori
for stimulating discussions related to this work.

\end{document}